\documentclass[aps,prb,preprint,superscriptaddress,showkeys]{revtex4-2}
\usepackage[T1]{fontenc}
\usepackage{times}
\usepackage{graphicx,color}
\usepackage{amsfonts,amsmath,amssymb,amsbsy}
\usepackage[colorlinks=true, linkcolor=blue, citecolor=blue, urlcolor=blue]{hyperref}
\renewcommand{\abstractname}{ABSTRACT}

\renewcommand{\figurename}{\textbf{Figure}}

\def\section#1{\medskip\noindent\textbf{#1}\par}
\makeatletter
\renewcommand{\fnum@figure}{\figurename~\textbf{\thefigure}}
\makeatother
\def\blue#1{\textcolor{blue}{#1}}
\begin{document}

\title{{\Large\sffamily Chiral Skyrmions Interacting with Chiral Flowers}}

\author{Xichao Zhang}
\thanks{X.Z. and J.X. contributed equally to this work.}
\affiliation{Department of Applied Physics, Waseda University, Okubo, Shinjuku-ku, Tokyo 169-8555, Japan}

\author{Jing Xia}
\thanks{X.Z. and J.X. contributed equally to this work.}
\affiliation{Department of Electrical and Computer Engineering, Shinshu University, 4-17-1 Wakasato, Nagano 380-8553, Japan}

\author{Oleg A. Tretiakov}
\affiliation{School of Physics, The University of New South Wales, Sydney 2052, Australia}

\author{Motohiko Ezawa}
\affiliation{Department of Applied Physics, The University of Tokyo, 7-3-1 Hongo, Tokyo 113-8656, Japan}

\author{Guoping Zhao}
\affiliation{College of Physics and Electronic Engineering, Sichuan Normal University, Chengdu 610068, China}

\author{Yan Zhou}
\affiliation{School of Science and Engineering, The Chinese University of Hong Kong, Shenzhen, Guangdong 518172, China}

\author{\\ Xiaoxi Liu}
\email[Email:~]{liu@cs.shinshu-u.ac.jp}
\affiliation{Department of Electrical and Computer Engineering, Shinshu University, 4-17-1 Wakasato, Nagano 380-8553, Japan}

\author{Masahito Mochizuki}
\email[Email:~]{masa_mochizuki@waseda.jp}
\affiliation{Department of Applied Physics, Waseda University, Okubo, Shinjuku-ku, Tokyo 169-8555, Japan}

\begin{abstract}
The chiral nature of active matter plays an important role in the dynamics of active matter interacting with chiral structures. Skyrmions are chiral objects, and their interactions with chiral nanostructures can lead to intriguing phenomena. Here, we explore the random-walk dynamics of a thermally activated chiral skyrmion interacting with a chiral flower-like obstacle in a ferromagnetic layer, which could create topology-dependent outcomes. It is a spontaneous mesoscopic order-from-disorder phenomenon driven by the thermal fluctuations and topological nature of skyrmions that exists only in ferromagnetic and ferrimagnetic systems. The interactions between the skyrmions and chiral flowers at finite temperatures can be utilized to control the skyrmion position and distribution without applying any external driving force or temperature gradient. The phenomenon that thermally activated skyrmions are dynamically coupled to chiral flowers may provide a new way to design topological sorting devices.
\end{abstract}

\date{December 7, 2023}

\preprint{\textsl{\href{https://doi.org/10.1021/acs.nanolett.3c03792}{DOI:~10.1021/acs.nanolett.3c03792}}}


\maketitle

\clearpage


Chiral skyrmions are versatile topological objects with fixed chirality in magnets with chiral exchange interactions~\cite{Bogdanov_1989,Roszler_NATURE2006,Nagaosa_NNANO2013,Mochizuki_Review,Fert_NATREVMAT2017,Everschor_JAP2018,Zhang_JPCM2020,Gobel_PP2021,Reichhardt_RMP2022,Del-Valle_2022,Wiesendanger_Review2016,Finocchio_JPD2016,Marrows_APL2021,Kanazawa_AM2017,Wanjun_PHYSREP2017}.
They can be created in magnetic thin films~\cite{Romming_SCIENCE2013,Heinze_NP2011}, multilayers~\cite{Wanjun_SCIENCE2015,Woo_NM2016,ML_NN2016}, and bulk nanostructures~\cite{Muhlbauer_SCIENCE2009,Yu_NATURE2010,Birch_NC2021}, where they can also be driven into motion by external forces~\cite{Nagaosa_NNANO2013,Mochizuki_Review,Fert_NATREVMAT2017,Everschor_JAP2018,Zhang_JPCM2020,Gobel_PP2021,Reichhardt_RMP2022,Del-Valle_2022,Wiesendanger_Review2016,Finocchio_JPD2016,Marrows_APL2021,Kanazawa_AM2017,Wanjun_PHYSREP2017}.
As skyrmions are usually rigid and nonvolatile~\cite{Nagaosa_NNANO2013,Mochizuki_Review,Fert_NATREVMAT2017,Everschor_JAP2018,Zhang_JPCM2020,Gobel_PP2021,Reichhardt_RMP2022,Del-Valle_2022,Wiesendanger_Review2016,Finocchio_JPD2016,Marrows_APL2021,Kanazawa_AM2017,Wanjun_PHYSREP2017,Lin_PRB2013}, they could be employed as nanoscale information carriers in next-generation information processing applications~\cite{Kang_PIEEE2016}, including data storage~\cite{Sampaio_NN2013,Tomasello_SREP2014} and logic computing~\cite{Xichao_SREP2015B}.
Recent studies also suggest that skyrmions can be used as building blocks in future non-conventional applications, including the neuromorphic~\cite{Song_NE2020} and quantum computing~\cite{Psaroudaki_PRL2021,Xia_PRL2022}.

The skyrmion dynamics is essential for skyrmionic devices.
The dynamics of chiral skyrmions stabilized by chiral exchange interactions~\cite{Dzyaloshinskii_1958,Moriya_1960} in ferromagnets include two aspects, i.e., the motion driven by applied forces~\cite{Nagaosa_NNANO2013,Mochizuki_Review,Fert_NATREVMAT2017,Everschor_JAP2018,Zhang_JPCM2020,Gobel_PP2021,Reichhardt_RMP2022,Del-Valle_2022,Wiesendanger_Review2016,Finocchio_JPD2016,Marrows_APL2021,Kanazawa_AM2017,Wanjun_PHYSREP2017,Kang_PIEEE2016} and the spontaneous diffusion induced by thermal fluctuations~\cite{Kong_PRL2013,Lin_PRL2014,Troncoso_AP2014,Tretiakov_PRL2016,Reichhardt_NJP2016,Miltat_PRB2018,Nozaki_APL2019,Reichhardt_JPCM2019,Zhao_PRL2020,Wang_NE2020,Yao_IEEE2020,Jing_PRB2021,Zhou_PRB2021,Miki_JPSJ2021,Suzuki_PLA2021,Ishikawa_APL2021,Yu_NC2021,Song_AFM2021,Kerber_PRA2021,Kong_PRB2021,Weissenhofer_PRL2021,Weissenhofer_PRB2023,Gruber_AM2023,Dohi_NC2023,Schutte_PRB2014}.
For example, a skyrmion driven by the spin-orbit torques may show the skyrmion Hall effect~\cite{Wanjun_NPHYS2017,Litzius_NPHYS2017,Reichhardt_JPCM2019,Reichhardt_NJP2016}, where the skyrmion moves at an angle with respect to the applied current direction.
On the other hand, a skyrmion driven by thermal effects may show the Brownian gyromotion~\cite{Troncoso_AP2014,Schutte_PRB2014,Tretiakov_PRL2016,Miltat_PRB2018,Nozaki_APL2019,Zhao_PRL2020,Zhou_PRB2021,Miki_JPSJ2021,Suzuki_PLA2021,Ishikawa_APL2021,Weissenhofer_PRB2023}, where the skyrmion tends to move in circular trajectories during the random walk.
Skyrmions can also be driven into directional motion by thermal gradients~\cite{Yu_NC2021,Wang_NE2020,Kong_PRL2013,Lin_PRL2014}.
Both the skyrmion Hall effect and skyrmion Brownian diffusion in the ferromagnetic and ferrimagnetic systems depend on the topological charge carried by the skyrmion (i.e., the skyrmion number), which is defined as $Q=\frac{1}{4\pi}\int\boldsymbol{m}\cdot(\frac{\partial\boldsymbol{m}}{\partial x}\times\frac{\partial\boldsymbol{m}}{\partial y})dxdy$ with $\boldsymbol{m}$ being the reduced net magnetization~\cite{Zhang_JPCM2020}.
The topology-dependent dynamic behaviors of skyrmions, either spontaneous or forced, are fundamental for practical applications and require precise control in nanostructures.

An important issue in the control and manipulation of skyrmion dynamics in nanostructures is the skyrmion-substrate interactions~\cite{Reichhardt_RMP2022,Reichhardt_PRB2015A,Reichhardt_NJP2015,Reichhardt_PRB2019,Vizarim_NJP2020,Zhang_Laminar2023}.
In active matter systems, the particle-particle and particle-substrate interactions play an important role in the particle dynamics~\cite{Bechinger_2016,Reichhardt_2017}.
As nanoscale skyrmions are usually rigid and can show self-motion at finite temperature (i.e., Brownian motion), they can also be treated as a special type of active quasiparticles and interact with the substrate effectively~\cite{Lin_PRB2013,Reichhardt_RMP2022,Reichhardt_PRB2015A,Reichhardt_NJP2015,Reichhardt_PRB2019,Vizarim_NJP2020,Li_SM2020,Zhang_Laminar2023}.
Moreover, the skyrmion-substrate interactions may result in some unique features due to the nontrivial topological nature of skyrmions~\cite{Lin_PRB2013,Reichhardt_RMP2022,Reichhardt_PRB2015A,Reichhardt_NJP2015,Reichhardt_PRB2019,Vizarim_NJP2020,Li_SM2020,Zhang_Laminar2023}.

In 2013, Mijalkov and Volpe demonstrated the possibility that particle-like chiral microswimmers performing circular active Brownian motion can be sorted in a chiral environment formed by using some static obstacle patterns on the substrate~\cite{Volpe_2013,Volpe_2014}, where the chirality of circular Brownian motion couples to chiral features present in the environment.
As skyrmions also show circular Brownian motion due to their nontrivial topology~\cite{Troncoso_AP2014,Schutte_PRB2014,Tretiakov_PRL2016,Miltat_PRB2018,Nozaki_APL2019,Zhao_PRL2020,Zhou_PRB2021,Miki_JPSJ2021,Suzuki_PLA2021,Ishikawa_APL2021,Weissenhofer_PRB2023}, it is therefore envisioned that the thermally activated random-walk dynamics of skyrmions may also be modified in a chiral environment due to the skyrmion-substrate interactions, which is the focus of this work.
However, it should be noted that active matter systems have some form of self-propulsion~\cite{Bechinger_2016,Reichhardt_2017,Volpe_2013,Volpe_2014}, while the Brownian skyrmions are only undergoing thermal motion and are not self-propelled.

\begin{figure}[h!]
\centerline{\includegraphics[width=0.75\textwidth]{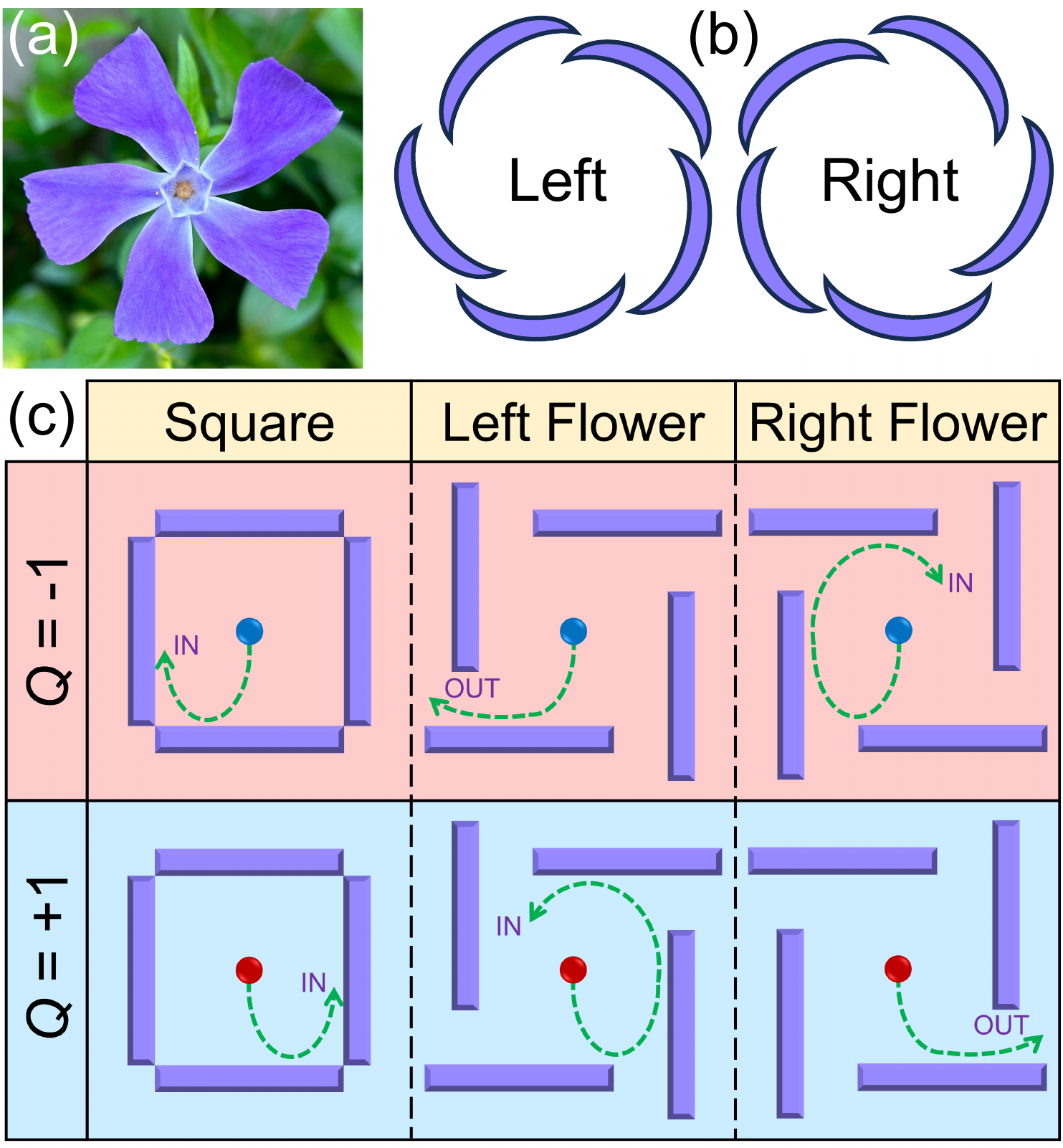}}
\caption{%
A thermally activated chiral skyrmion interacting with a chiral flower-like obstacle.
(\textbf{a}) An exemplary chiral flower (\textit{Vinca minor}) found in Shinjuku City by the authors, which shows a fixed left-contort corolla.
(\textbf{b}) Diagrams showing the left-right asymmetry in flowers with fixed corolla contortion, i.e., the left-contort and right-contort corollas.
(\textbf{c}) Typical desired outcomes of a chiral skyrmion interacting with a chiral flower or a square: skyrmion confined (i.e., ``IN'') and skyrmion escaped (i.e, ``OUT''). The outcomes depend on the skyrmion number $Q$ as well as the left-right asymmetry of the chiral flower. The skyrmions with $Q=-1$ and $Q=+1$ are denoted by blue and red dots, respectively.
}
\label{FIG1}
\end{figure}

\begin{figure*}[t]
\centerline{\includegraphics[width=0.99\textwidth]{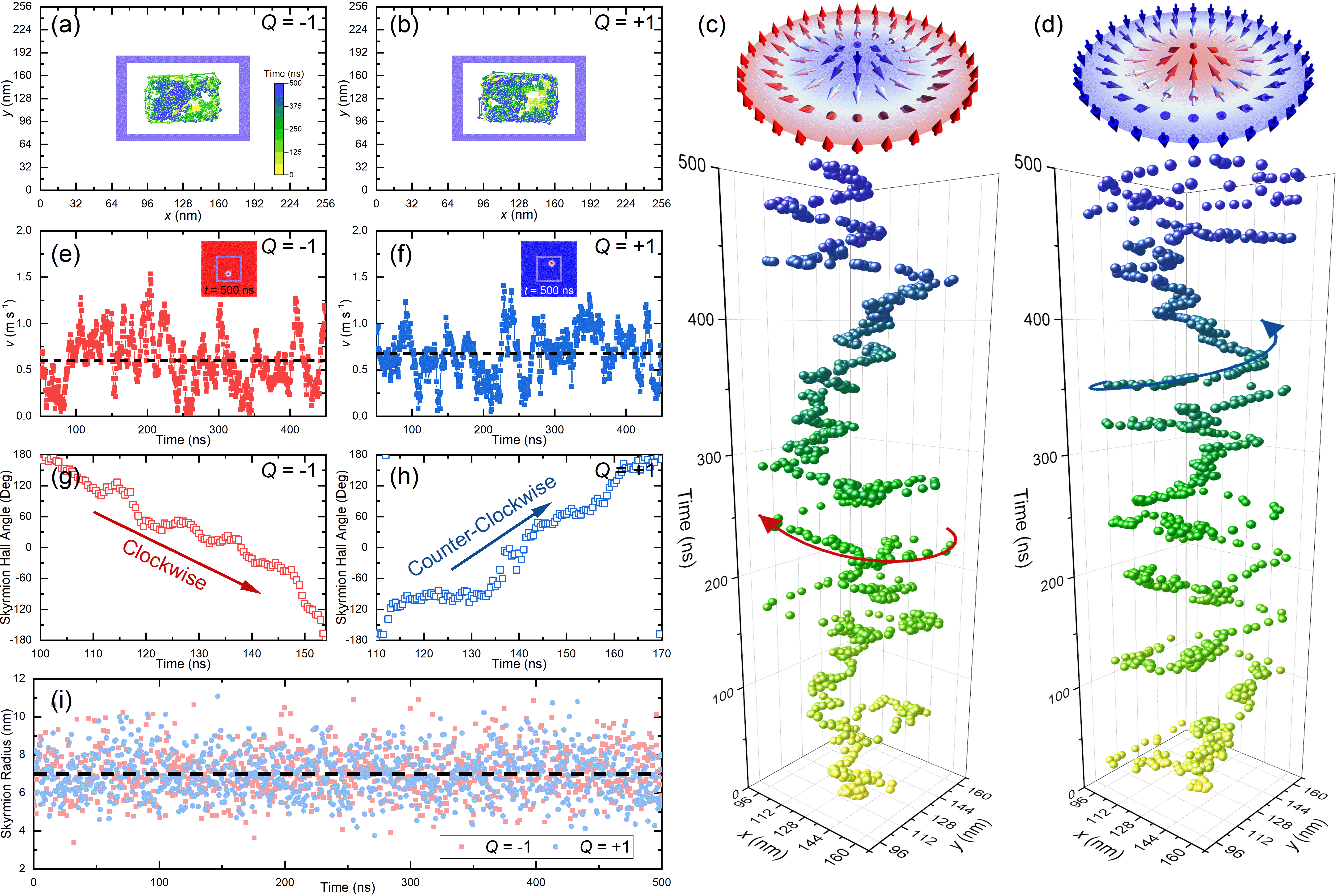}}
\caption{%
A thermally activated skyrmion confined by a square obstacle.
Typical trajectories of (\textbf{a}) a skyrmion with $Q=-1$ and (\textbf{b}) a skyrmion with $Q=+1$ confined by a square are given.
Three-dimensional illustrations show the time-dependent skyrmion position and the skyrmion texture with (\textbf{c}) $Q=-1$ or (\textbf{d}) $Q=+1$. The skyrmions with $Q=-1$ and $Q=+1$ show clockwise and counterclockwise circular motion along the inner edges of the square obstacle, respectively.
Time-dependent velocities $v$ of the skyrmions with (\textbf{e}) $Q=-1$ and (\textbf{f}) $Q=+1$ are given.
Time-dependent skyrmion Hall angles of the skyrmions with (\textbf{g}) $Q=-1$ and (\textbf{h}) $Q=+1$ are also shown for selected time ranges, indicating the clockwise and counterclockwise circular motion, respectively.
(\textbf{i}) Time-dependent skyrmion radius.
The skyrmion dynamics is simulated at $T=150$ K for $500$ ns with a time step of $0.5$ ns.
The time step is small enough to show the Brownian motion with a reasonable precision (see Figure~\blue{S1} in the Supporting Information).
}
\label{FIG2}
\end{figure*}

\begin{figure*}[t]
\centerline{\includegraphics[width=0.99\textwidth]{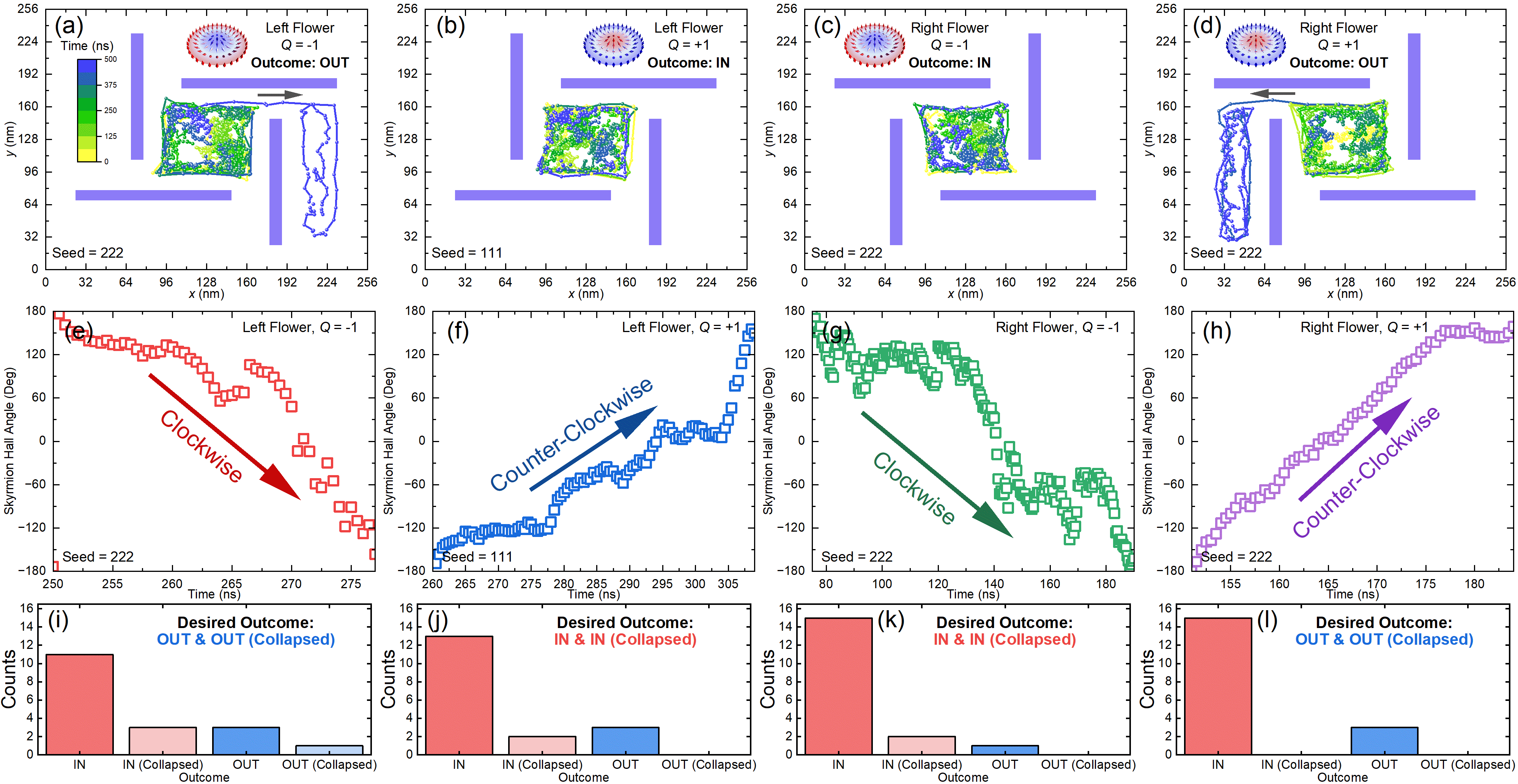}}
\caption{%
A thermally activated skyrmion interacting with a left-handed or right-handed flower.
(\textbf{a}) Typical trajectory of a skyrmion with $Q=-1$ interacting with a left-handed flower. The skyrmion escapes from the left-handed flower due to its clockwise Brownian gyromotion, i.e, the outcome is a desired ``OUT''.
(\textbf{b}) Typical trajectory of a skyrmion with $Q=+1$ interacting with a left-handed flower. The skyrmion is confined by the left-handed flower due to its counterclockwise Brownian gyromotion, i.e, the outcome is a desired ``IN''.
(\textbf{c}) Typical trajectory of a skyrmion with $Q=-1$ interacting with a right-handed flower. The skyrmion is confined by the right-handed flower due to its clockwise Brownian gyromotion, i.e, the outcome is a desired ``IN''.
(\textbf{d}) Typical trajectory of a skyrmion with $Q=+1$ interacting with a right-handed flower. The skyrmion escapes from the right-handed flower due to its counterclockwise Brownian gyromotion, i.e, the outcome is a desired ``OUT''.
(\textbf{e})-(\textbf{h}) Time-dependent skyrmion Hall angles for selected time ranges, corresponding to (\textbf{a}) and (\textbf{b}), respectively.
(\textbf{i})-(\textbf{l}) The outcome counts for a skyrmion with $Q=\pm 1$ interacting with a left- or right-handed flower.
Eighteen simulations are done with different random seeds for each skyrmion-flower configuration.
Both the desired and undesired outcomes are counted.
The skyrmion dynamics is simulated at $T=150$ K for $500$ ns with a time step of $0.5$ ns.
}
\label{FIG3}
\end{figure*}

\begin{figure}[t]
\centerline{\includegraphics[width=0.75\textwidth]{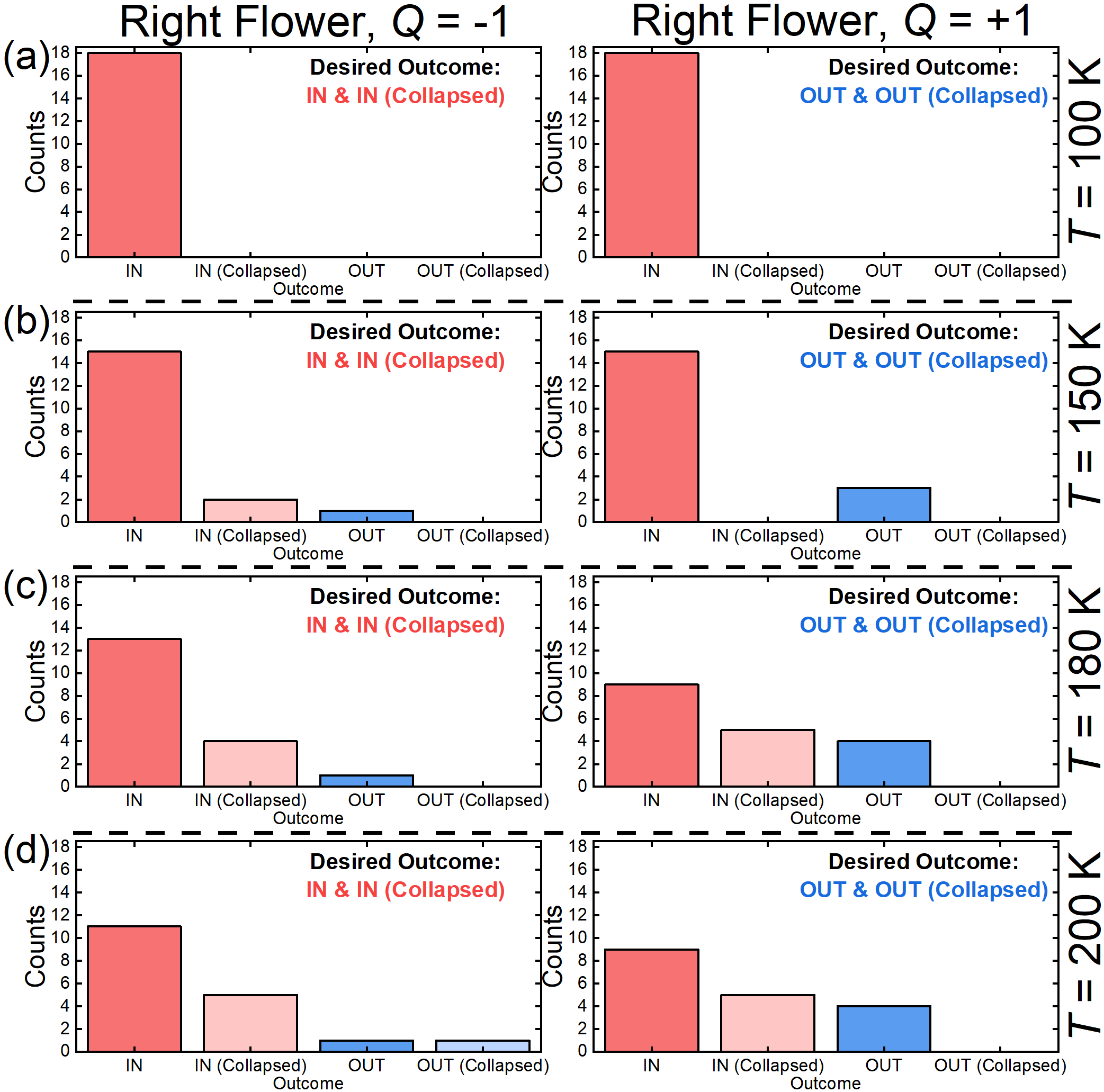}}
\caption{%
Outcome counts for a skyrmion interacting with a right-handed flower at different temperatures.
A skyrmion with $Q=\pm 1$ interacting with a right-handed flower at (\textbf{a}) $T=100$ K, (\textbf{b}) $T=150$ K, (\textbf{c}) $T=180$ K, and (\textbf{d}) $T=200$ K.
Eighteen simulations are done with different random seeds for each temperature.
Both desired and undesired outcomes are counted.
The skyrmion dynamics is simulated for $500$ ns with a time step of $0.5$ ns.
}
\label{FIG4}
\end{figure}


The square and chiral flower-like obstacles considered in this work are schematically depicted in Figure~\ref{FIG1}.
A real example of a chiral flower is given in Figure~\ref{FIG1}(a), which is a left-handed flower showing a fixed left-contort corolla.
The chirality of a flower, either left-handed or right-handed [Figure~\ref{FIG1}(b)], is an important property of floral symmetry. Some flowers have a contort petal aestivation, which is most pronounced in floral buds and may be less prominent in open flowers~\cite{Endress_1999,Endress_2001}. In chiral flowers, two morphs are possible as shown in Figure~\ref{FIG1}(b): contorted to the left and contorted to the right~\cite{Endress_1999,Endress_2001}.
Micro- and nanostructures mimicking chiral flowers may control the dynamics of active chiral matter~\cite{Volpe_2013,Volpe_2014,Bechinger_2016,Reichhardt_2017} as well as thermally activated chiral skyrmions.
In Figure~\ref{FIG1}(c), a thermally activated skyrmion shows clockwise or counterclockwise Brownian gyromotion, which depends on the sign of $Q$. If a skyrmion is initially placed within a chiral flower, then it may escape from or be confined by the chiral flower.
The outcome depends on the chirality of the flower and the sign of $Q$, which could result in the topological sorting and create an order (sorting)-from-disorder (Brownian motion) phenomenon.
However, if a skyrmion is initially placed within a square obstacle, then it will be confined by the square.



We first show the typical Brownian gyromotion of a ferromagnetic skyrmion within a square obstacle in a two-dimensional model, which results in the confinement of the skyrmion~\cite{Muratov_2022,Ishikawa_APL2021,Miki_JPSJ2021}.
The length, width, and thickness of the ferromagnetic layer equal $256$, $256$, and $1$ nm, respectively.
The square pattern is made of four obstacle bars, which are rectangle regions locally modified to have enhanced PMA $K_{\text{o}}$. We assume that $K_{\text{o}}/K=10$ in order to make sure that the skyrmions cannot penetrate the obstacle boundary~\cite{Xichao_PRB2022B}.
Such a square pattern on the ferromagnetic substrate can, in principle, be fabricated in experiments~\cite{Juge_NL2021,Ohara_NL2021,Zhang_CP2021,Ishikawa_APL2021}.
The width of each obstacle bar is $10$ nm, and the distance between two parallel inner edges of the square pattern is $100$ nm. The distance between the two parallel outer edges is thus $120$ nm. The square center overlaps the ferromagnetic layer center, as indicated in Figure~\ref{FIG2}(a).
Other modeling details and parameters are given in Methods.

Initially, a skyrmion is placed and relaxed at the center of the ferromagnetic layer. We then simulate the thermal random-walk dynamics (i.e., the Brownian motion) of the skyrmion at a temperature of $T=150$ K for $500$ ns.
The trajectories of the skyrmions with $Q=-1$ and $Q=+1$ are given in Figures~\ref{FIG2}(a) and~\ref{FIG2}(b), respectively. The skyrmion with $Q=-1$ shows clockwise Brownian gyromotion [Figure~\ref{FIG2}(c)] (see Movie S1), while the skyrmion with $Q=+1$ shows counterclockwise Brownian gyromotion [Figure~\ref{FIG2}(d)] (see Movie S2).
The skyrmions may move along the inner edges of the square and follow the direction of the intrinsic Brownian gyromotion. The confinement leads to a square shape of the overlapped skyrmion position distribution for $500$ ns of simulation.
The skyrmion motion guided by the square edges is similar to that guided by grain boundaries~\cite{Zhou_PRB2021}, which may enhance the skyrmion diffusion.
The Brownian gyromotion of a skyrmion is a feature of its topological nature, which is due to the Magnus force associated with the net skyrmion number~\cite{Troncoso_AP2014,Schutte_PRB2014,Tretiakov_PRL2016,Miltat_PRB2018,Nozaki_APL2019,Zhao_PRL2020,Zhou_PRB2021,Miki_JPSJ2021,Suzuki_PLA2021,Ishikawa_APL2021,Weissenhofer_PRB2023}. We note that the Magnus force is absent in the antiferromagnetic system, where a skyrmion may not show Brownian gyromotion~\cite{Weissenhofer_PRB2023,Tretiakov_PRL2016}.

The time-dependent velocities of the skyrmions with $Q=-1$ and $Q=+1$ interacting with the square are given in Figure~\ref{FIG2}(e) and~\ref{FIG2}(f), respectively. The velocity is randomly fluctuating with time and the mean value for the skyrmion with $Q=-1$ equals $0.60$ m s$^{-1}$, which is almost the same with that of the skyrmion with $Q=-1$ ($0.68$ m s$^{-1}$). 
The clockwise and counterclockwise Brownian gyromotion can also be seen from the time-dependent skyrmion Hall angle defined as $\theta_{\text{SkHE}}=\text{arctan}(v_y/v_x)$ with $v_x$ and $v_y$ being the $x$ and $y$ components of the skyrmion velocity, respectively.
In Figure~\ref{FIG2}(g), the clockwise skyrmion motion is indicated by the continuous decrease and sharp increase of $\theta_{\text{SkHE}}(t)$. The counterclockwise skyrmion motion is indicated by the continuous increase and sharp decrease of $\theta_{\text{SkHE}}(t)$ [Figure~\ref{FIG2}(h)].
The radius of the skyrmion with $Q=\pm 1$ interacting with the square is also fluctuating with time, and its mean value equals $7$ nm during $500$ ns of simulation [Figure~\ref{FIG2}(i)].


We also demonstrate two typical outcomes of a thermally activated skyrmion with $Q=\pm 1$ interacting with a left- or right-handed flower-like obstacle pattern in a two-dimensional model.
The chiral flower pattern is made of four obstacle bars, which are rectangle regions with enhanced PMA $K_{\text{o}}/K=10$.
The width and length of each obstacle bar are equal to $10$ and $120$ nm, respectively. The distance between the two parallel inner edges of the obstacle bars is $100$ nm. The opening width between the two orthogonal obstacle bars is set to $30$ nm.
The chiral flower center overlaps the ferromagnetic layer center, as shown in Figure~\ref{FIG3}(a).
The opening width should be wider but not much wider than the skyrmion diameter, and the area within the chiral flower should not be too large. Otherwise, the skyrmion may not interact with the chiral flower in an effective way depending on its diffusion at a given temperature.
If the opening width is much larger than the skyrmion diameter, the skyrmion should be easier to escape and enter the flower, and travel in all possible directions equally often during long times, leading to achiral results.

A skyrmion is initially placed and relaxed at the ferromagnetic layer center. We then simulate the thermal random-walk dynamics of a skyrmion at a given temperature for $500$ ns.
We first show typical desired outcomes of the skyrmion with $Q=\pm 1$ interacting with a left- or right-handed flower at $T=150$ K in Figures~\ref{FIG3}(a)-\ref{FIG3}(d).
The skyrmion with $Q=-1$ shows the clockwise Brownian gyromotion [Figures~\ref{FIG3}(e) and~\ref{FIG3}(g)]. Consequently, its interactions with the left-handed [Figure~\ref{FIG3}(a)] and right-handed [Figure~\ref{FIG3}(c)] chiral flowers lead to the desired outcomes ``OUT'' and ``IN'', respectively, within $500$ ns of simulation (see Movie S3 and Movie S4).
The skyrmions with $Q=+1$ showing the counterclockwise Brownian gyromotion [Figures~\ref{FIG3}(f) and~\ref{FIG3}(h)] and interacting with the left-handed [Figure~\ref{FIG3}(b)] and right-handed [Figure~\ref{FIG3}(d)] chiral flowers show the desired outcomes ``IN'' and ``OUT'', respectively (see Movie S5 and Movie S6).
The desired outcomes for the four skyrmion-flower configurations are summarized schematically in Figure~\ref{FIG1}(c).
The time-dependent velocity and skyrmion radius during the skyrmion-flower interaction are given in Figure~\blue{S2} (see Supporting Information),
and the time-dependent total energy of the system is given in Figure~\blue{S3} (see Supporting Information).

The desired outcomes may be achieved when the skyrmion interacts effectively with the chiral flower for a long enough time. However, as a skyrmion has a certain lifetime at finite temperature, it may collapse before or after achievement of the desired outcomes.
With reasonable computational workload, we carry out $18$ simulations with different random seeds for each temperature and skyrmion-flower configuration and summarize both the desired and undesired outcomes.
For the skyrmion with $Q=-1$ interacting with a left-handed flower, we obtain four events of desired outcome in $18$ simulations [Figure~\ref{FIG3}(i)].
For the skyrmion with $Q=+1$ interacting with a left-handed flower, we obtain $15$ events of desired outcome [Figure~\ref{FIG3}(j)], including two cases in which the skyrmion collapses within the chiral flower.
We also note that three undesired ``OUT'' events happen, which may be due to the fact that the skyrmion size is transiently much smaller than the opening width when it moves to an exit of the left-handed flower along the inner edge of the obstacle bar (see Movie S7).
Such a situation may be avoided by slightly reducing the opening width or increasing the skyrmion size; however, it also indicates that the skyrmion is able to travel along the path unfavored by the skyrmion-flower interaction, especially during long times or when the skyrmion-flower interaction is ineffective.
For the skyrmion with $Q=-1$ interacting with a right-handed flower, we obtain $17$ events of the desired outcome [Figure~\ref{FIG3}(k)].
For the skyrmion with $Q=+1$ interacting with a right-handed flower, we obtain three events of desired outcome [Figure~\ref{FIG3}(l)].
We note that when the desired outcome is the ``IN'' event, both effective and ineffective skyrmion-flower interactions may result in the desired outcome.
We also show the skyrmion interacting with an achiral square with corner gaps in the Supporting Information (see Figure~\blue{S4}), where it is expected that both skyrmions with $Q=\pm 1$ can escape easily to explore the whole sample, and the outcomes are independent of $Q$.

In Figure~\ref{FIG4}, we further show that the counts of achieving the desired and undesired outcomes within the $500$ ns-long simulation of the skyrmion-flower interaction depend on the temperature.
When the temperature is too low [Figure~\ref{FIG4}(a); $T=100$ K], the skyrmion diffusion is weak and it cannot interact with the chiral flower effectively.
In such a case, we obtain $18$ ``IN'' events in $18$ simulations for the skyrmions with $Q=\pm 1$ within a right-handed flower.
When $T=150$ K [Figure~\ref{FIG4}(b)], we obtain $17$ desired ``IN'' events for the skyrmion with $Q=-1$ interacting a right-handed flower, and three desired ``OUT'' events for the skyrmion with $Q=+1$ interacting with a right-handed flower.
When $T=180$ K [Figure~\ref{FIG4}(c)], we obtain $17$ desired ``IN'' events for the skyrmion with $Q=-1$ interacting a right-handed flower, and four desired ``OUT'' events for the skyrmion with $Q=+1$ interacting with a right-handed flower.
It suggests that the skyrmion-flower interaction could be more effective due to more active skyrmion at elevated temperature.
However, when the temperature is too high [Figure~\ref{FIG4}(d); $T=200$ K], the thermal fluctuations may result in the collapse of the skyrmion in more simulations due to the significantly reduced skyrmion lifetime.
The mean skyrmion size may also increase with the temperature, while the size and geometry of the chiral flower are fixed.
Therefore, the skyrmion may not interact with the chiral flower effectively when the temperature is too high.


In conclusion, we have studied the thermal random-walk dynamics of a ferromagnetic skyrmion in a chiral environment, where the interactions between skyrmions and chiral obstacles (i.e., the chiral flowers) could lead to topology-dependent spontaneous sorting of skyrmions.
The position of a skyrmion can be manipulated by using a simple chiral flower-like obstacle pattern at finite temperature in the absence of an external drive if the skyrmion is placed initially at the center of the flower, which is a state of artificially low entropy.
Namely, an effective interaction between the chiral flower (either left or right) and the skyrmion (either $Q = -1$ or $+1$) could result in the escape or confinement of the skyrmion. For both outcomes, as the skyrmion tends to explore the whole space (i.e., inside and outside the flower) during long times due to its thermal diffusion, the disorder and entropy of the system should increase with time, while the total energy is conserved over time despite fluctuations due to the thermal effect. Thus, the skyrmion behaviors in such a closed system are in line with the first and second laws of thermodynamics. However, we point out that some systems are more ordered even when there is increased entropy in the ordered state~\cite{Geng_SA2019}.

Our results reveal the unique thermal dynamics of chiral topological spin textures interacting with chiral structures.
Our results also suggest that it is possible to build a topological sorting device based on chiral flower-like structures, in which skyrmions with opposite signs of topological charges could generate different dynamic outcomes.


\vbox{}
\noindent\textbf{\sffamily METHODS}

\noindent
\textbf{\sffamily Computational Simulations.}
The simulations are performed by using the micromagnetic simulator \textsc{mumax$^3$}~\cite{MuMax1,MuMax2} on several commercial graphics processing units, including NVIDIA GeForce RTX 3070 and RTX 3060 Ti.
The magnetization dynamics at finite temperature is governed by the stochastic Landau-Lifshitz-Gilbert (LLG) equation~\cite{MuMax1,MuMax2},
\begin{equation}
\partial_{t}\boldsymbol{m}=-\gamma_{0}\boldsymbol{m}\times(\boldsymbol{h}_{\text{eff}}+\boldsymbol{h}_{\text{f}})+\alpha(\boldsymbol{m}\times\partial_{t}\boldsymbol{m}),
\label{eq:LLGS-CPP}
\end{equation}
where $\boldsymbol{m}=\boldsymbol{M}/M_{\text{S}}=1$ is the reduced magnetization,
$M_{\text{S}}$ is the saturation magnetization,
$t$ is the time,
$\gamma_0$ is the absolute gyromagnetic ratio,
$\alpha$ is the Gilbert damping parameter,
$\boldsymbol{h}_{\rm{eff}}=-\frac{1}{\mu_{0}M_{\text{S}}}\cdot\frac{\delta\varepsilon}{\delta\boldsymbol{m}}$ is the effective field with $\mu_{0}$ and $\varepsilon$ being the vacuum permeability constant and average energy density, respectively.
$\boldsymbol{h}_{\text{f}}$ is a thermal fluctuating field satisfying~\cite{MuMax1,MuMax2}
\begin{align}
&<h_{i}(\boldsymbol{x},t)>=0, \\ \notag
&<h_{i}(\boldsymbol{x},t)h_{j}(\boldsymbol{x}',t')>=\frac{2\alpha k_{\text{B}}T}{M_\text{S}\gamma_0\mu_0 V}\delta_{ij}\delta(\boldsymbol{x}-\boldsymbol{x}')\delta(t-t'),
\label{eq:correlations}
\end{align}
where $i$ and $j$ are Cartesian components, $k_{\text{B}}$ is the Boltzmann constant, $T$ is the temperature, and $V$ is the volume of a single mesh cell.
$\delta_{ij}$ and $\delta(\dots)$ denote the Kronecker and Dirac delta symbols, respectively.
The energy terms considered in the model include the ferromagnetic exchange energy, interface-induced chiral exchange energy, perpendicular magnetic anisotropy (PMA) energy, and demagnetization energy.
Thus, the average energy density is given as~\cite{MuMax1,MuMax2}
\begin{equation}
\label{eq:energy-density}
\begin{split}
\varepsilon=&A\left(\nabla\boldsymbol{m}\right)^{2}+D\left[m_{z}\left(\boldsymbol{m}\cdot\nabla\right)-\left(\nabla\cdot\boldsymbol{m}\right)m_{z}\right] \\
-&K(\boldsymbol{n}\cdot\boldsymbol{m})^2-\frac{M_{\text{S}}}{2}(\boldsymbol{m}\cdot\boldsymbol{B}_{\text{d}}),
\end{split}
\end{equation}
where $A$, $D$, and $K$ are the ferromagnetic exchange, DM interaction, and PMA constants, respectively.
$\boldsymbol{B}_{\text{d}}$ is the demagnetization field. $\boldsymbol{n}$ is the unit surface normal vector. $m_z$ is the out-of-plane component of $\boldsymbol{m}$.
The default magnetic parameters are~\onlinecite{Sampaio_NN2013,Tomasello_SREP2014,Xichao_SREP2015B,Xichao_PRB2022B}:
$\gamma_{0}=2.211\times 10^{5}$ m A$^{-1}$ s$^{-1}$,
$\alpha=0.1$,
$M_{\text{S}}=580$ kA m$^{-1}$,
$A=15$ pJ m$^{-1}$,
$K=0.8$ MJ m$^{-3}$,
and $D=3$ mJ m$^{-2}$.
The mesh size is $2$ $\times$ $2$ $\times$ $1$ nm$^3$, which ensures good computational accuracy and efficiency.
The finite-temperature simulation is performed with a fixed integration time step of $10$ fs and a given random seed.

\vbox{}
\noindent\textbf{\sffamily ASSOCIATED CONTENT}

\noindent
\textbf{\sffamily Supporting Information}

\noindent
The Supporting Information is available free of charge at [\href{https://doi.org/10.1021/acs.nanolett.3c03792}{https://doi.org/10.1021/acs.nanolett.-3c03792}].

\vbox{}\noindent
Additional simulation results, including the time-dependent velocity and radius of a skyrmion interacting with a left or right chiral flower, the time-dependent total energy of the system, the interaction between a skyrmion with a square with corner gaps, and the time-step-dependent skyrmion trajectories. (PDF)

\vbox{}\noindent
Movie S1: A thermally activated skyrmion with $Q=-1$ interacting with a square obstacle pattern. The skyrmion dynamics is simulated at $T=150$ K for $500$ ns with a time step of $0.5$ ns. The random seed equals $111$. It shows the confinement of the skyrmion by the square. In order to show both the skyrmion and obstacle pattern, the movie shows the time-dependent magnetic anisotropy energy density instead of the magnetization. (MP4)

\vbox{}\noindent
Movie S2: A thermally activated skyrmion with $Q=+1$ interacting with a square obstacle pattern. The skyrmion dynamics is simulated at $T=150$ K for $500$ ns with a time step of $0.5$ ns. The random seed equals $111$. It shows the confinement of the skyrmion by the square. In order to show both the skyrmion and obstacle pattern, the movie shows the time-dependent magnetic anisotropy energy density instead of the magnetization. (MP4)

\vbox{}\noindent
Movie S3: A thermally activated skyrmion with $Q=-1$ interacting with a left flower-like obstacle pattern. The skyrmion dynamics is simulated at $T=150$ K for $500$ ns with a time step of $0.5$ ns. The random seed equals $222$. The skyrmion escapes from the left flower as a desired outcome. In order to show both the skyrmion and obstacle pattern, the movie shows the time-dependent magnetic anisotropy energy density instead of the magnetization. (MP4)

\vbox{}\noindent
Movie S4: A thermally activated skyrmion with $Q=+1$ interacting with a left flower-like obstacle pattern. The skyrmion dynamics is simulated at $T=150$ K for $500$ ns with a time step of $0.5$ ns. The random seed equals $111$. The skyrmion is confined by the left flower as a desired outcome. In order to show both the skyrmion and obstacle pattern, the movie shows the time-dependent magnetic anisotropy energy density instead of the magnetization. (MP4)

\vbox{}\noindent
Movie S5: A thermally activated skyrmion with $Q=-1$ interacting with a right flower-like obstacle pattern. The skyrmion dynamics is simulated at $T=150$ K for $500$ ns with a time step of $0.5$ ns. The random seed equals $222$. The skyrmion is confined by the right flower as a desired outcome. In order to show both the skyrmion and obstacle pattern, the movie shows the time-dependent magnetic anisotropy energy density instead of the magnetization. (MP4)

\vbox{}\noindent
Movie S6: A thermally activated skyrmion with $Q=+1$ interacting with a right flower-like obstacle pattern. The skyrmion dynamics is simulated at $T=150$ K for $500$ ns with a time step of $0.5$ ns. The random seed equals $222$. The skyrmion escapes from the right flower as a desired outcome. In order to show both the skyrmion and obstacle pattern, the movie shows the time-dependent magnetic anisotropy energy density instead of the magnetization. (MP4)

\vbox{}\noindent
Movie S7: A thermally activated skyrmion with $Q=+1$ interacting with a left flower-like obstacle pattern. The skyrmion dynamics is simulated at $T=150$ K for $500$ ns with a time step of $0.5$ ns. The random seed equals $777$. The skyrmion escapes from the left flower as an undesired outcome. In order to show both the skyrmion and obstacle pattern, the movie shows the time-dependent magnetic anisotropy energy density instead of the magnetization. (MP4)

\vbox{}
\noindent\textbf{\sffamily AUTHOR INFORMATION}

\noindent
\textbf{\sffamily Corresponding Authors}

\noindent
Xiaoxi Liu; Email: liu@cs.shinshu-u.ac.jp

\noindent
Masahito Mochizuki; Email: masa$\_$mochizuki@waseda.jp

\vbox{}\noindent
Complete contact information is available at: [\href{https://doi.org/10.1021/acs.nanolett.3c03792}{https://doi.org/10.1021/acs.nanolett.3c03792}].

\vbox{}
\noindent\textbf{\sffamily Author Contributions}

\noindent
X.Z. and J.X. contributed equally to this work. X.Z., M.M. and X.L. conceived the idea. X.L. and M.M. coordinated the project. X.Z. and J.X. performed the computational simulation and the theoretical analysis. X.Z. took a picture of the chiral flower. X.Z. and J.X. drafted the paper and revised it with input from O.A.T., M.E., G.Z., Y.Z., X.L., and M.M. All authors discussed the results and reviewed the paper.

\vbox{}
\noindent\textbf{\sffamily Notes}

\noindent
The authors declare no competing financial interest.

\vbox{}
\noindent\textbf{\sffamily ACKNOWLEDGEMENTS}

\noindent
X.Z. and M.M. acknowledge support by CREST, the Japan Science and Technology Agency (Grant No. JPMJCR20T1).
M.M. also acknowledges support by the Grants-in-Aid for Scientific Research from JSPS KAKENHI (Grants No. JP20H00337 and No. JP23H04522), and the Waseda University Grant for Special Research Projects (Grant No. 2023C-140).
J.X. was a JSPS International Research Fellow supported by JSPS KAKENHI (Grant No. JP22F22061).
O.A.T. acknowledges support by the Australian Research Council (Grant No. DP200101027), the Cooperative Research Project Program at the Research Institute of Electrical Communication, Tohoku University (Japan), and by the NCMAS grant.
M.E. acknowledges support by CREST, JST (Grant No. JPMJCR20T2).
G.Z. acknowledges support by the National Natural Science Foundation of China (Grants No. 51771127, No. 51571126, and No. 51772004), and Central Government Funds of Guiding Local Scientific and Technological Development for Sichuan Province (Grant No. 2021ZYD0025).
Y.Z. acknowledges support by the National Natural Science Foundation of China (Grants No. 11974298 and No. 12374123), the Shenzhen Fundamental Research Fund (Grant No. JCYJ20210324120213037), the Shenzhen Peacock Group Plan (Grant No. KQTD20180413181702403), and the Guangdong Basic and Applied Basic Research Foundation (Grant No. 2021B1515120047).
X.L. acknowledges support by the Grants-in-Aid for Scientific Research from JSPS KAKENHI (Grants No. JP20F20363, No. JP21H01364, No. JP21K18872, and No. JP22F22061).

\vbox{}



\end{document}